\documentclass{ws-procs9x6-cpt22}

\usepackage{tikz}
\usetikzlibrary{positioning,fit,shapes.geometric,backgrounds,arrows,shapes}

\def\thpr{these proceedings}

\def\frac#1#2{{\textstyle{{#1}\over {#2}}}}

\def\ibar{{\mathrel{\rlap{$I$} \raise1pt\hbox{--}}}}

\def\lsim{\mathrel{\rlap{\lower4pt\hbox{\hskip1pt$\sim$}}
    \raise1pt\hbox{$<$}}}
\def\gsim{\mathrel{\rlap{\lower4pt\hbox{\hskip1pt$\sim$}}
    \raise1pt\hbox{$>$}}}
\def\sqr#1#2{{\vcenter{\vbox{\hrule height.#2pt
         \hbox{\vrule width.#2pt height#1pt \kern#1pt
         \vrule width.#2pt}
         \hrule height.#2pt}}}}

\def\etal{{\it et al.}}

\newcommand{\beq}{\begin{equation*}}
\newcommand{\eeq}{\end{equation*}}
\newcommand{\bea}{\begin{eqnarray}}
\newcommand{\eea}{\end{eqnarray}}
\newcommand{\bit}{\begin{itemize}}
\newcommand{\eit}{\end{itemize}}
\newcommand{\bcn}{\begin{center}}
\newcommand{\ecn}{\end{center}}

\begin{document}

\newcommand{\refeq}[1]{(\ref{#1})}
\def\etal {{\it et al.}}

\title{SME Gravity: Structure and Progress}

\author{Jay D.\ Tasson}

\address{
  Physics and Astronomy Department, Carleton College,\\
  Northfield, MN 55057, USA\\
{\rm LIGO-P2200209-v1}
}

\begin{abstract}
This proceedings contribution outlines the current structure of the gravity sector of the Standard-Model Extension
and summaries recent progress in gravitational wave analysis.
\end{abstract}

\bodymatter

\section{Lorentz violation in gravity}
The gravitational Standard-Model Extension (SME) \cite{SME1,SME2,SME3}
provides a field theoretic test framework
for Lorentz symmetry.
Originally motivated by the search for new physics at the Planck Scale,\cite{ks}
the search for Lorentz violation using the SME continues to be an active and growing area of research
several decades later,
as illustrated by the scope of both these proceedings 
and the
{\it Data Tables for Lorentz and CPT Violation}.\cite{datatables}

\begin{figure}
\begin{center}
\begin{tikzpicture}[%
    node distance = .4cm,
  inner sep=1mm,
  expt/.style={rectangle, trapezium left angle=120, trapezium right angle=60, thin, fill=gray!30, draw=black, align=center},
  action/.style={rectangle, trapezium left angle=120, trapezium right angle=60, thick, fill=gray!10, draw=black, align=center},
  oact/.style={rectangle, trapezium left angle=120, trapezium right angle=60, dashed, fill=gray!20, draw=black, align=center},
    loop/.style={ellipse, thick, fill=yellow!30, draw=yellow!50, align=center},
  title/.style={font=\LARGE\scshape,node distance=16pt, text=black!40, inner sep=1mm},
  background/.style={rectangle, rounded corners, fill=black!5, draw=black!15, inner sep=4mm}
]

\node[action] (akgrav)
       {minimal gravitational \\
         SME \cite{SME2}};
  \node[action] (lvpn) [above right = -0.5cm and 0.5cm of akgrav] {linearized\\
 pure-gravity \cite{lvpn}};
  \node[action] (lvgap) [ below =of lvpn] {linearized\\
 matter-gravity \cite{lvgap}};
  \node[expt] (exp) [above right = -0.7cm and 0.5cm of lvgap] {
        additional\\ phenomenology\cite{phenom,adv}\\
\&
        tests\cite{datatables}};
  \node[action] (akmm) [ below =1.5cm of akgrav] {complete linearized\\
         pure-gravity \cite{mkgrav}};
  \node[expt] (nonmin) [ right= 1cm of akmm] {additional\\ phenomenology \cite{bkx}\\
\&
        tests \cite{datatables,srexpt,psrnm,birf,zwang21,niu}};
  \node[action] (nonlin) [ below  = of akmm] {nonminimal \\ nonlinear gravity \cite{qb16}};

  \node[expt] (nlx) [ right= 1cm of nonlin] {tests \cite{datatables}};

  \node[oact] (other) [below right = -0.8cm and 0.5cm of nonmin] {models \\ \& theory \cite{rb}};

  \node[oact] (fin) [below = of other] {Finsler geometry\cite{be}};
  
  \node[fit=(akgrav)] (chart) {};
  
  \draw[->] (akgrav.east) -- (lvpn);
  \draw[->] (akgrav.east) -- (lvgap);
  \draw[->] (lvpn.east) -- (exp);
  \draw[->] (lvgap.east) -- (exp);
  \draw[<->] (akgrav.south) -- (akmm);
  \draw[->] (akmm.east) -- (nonmin);
  \draw[->] (nonlin.east) -- (nlx);
  
\end{tikzpicture}
\end{center}
\caption{Structure of the gravity sector as of CPT'19.
Light gray boxes show the various limits of this sector 
that had been explored to this point.
Work that builds out the search in the respective limits
appears in dark gray boxes.
Theoretical contributions are shown in dashed boxes.
While the structure shown is the same as seen\cite{jtcpt19} in CPT'19,
references have been updated to reflect the additional work done
on a number of nodes.} 
\label{asof19}
\end{figure}
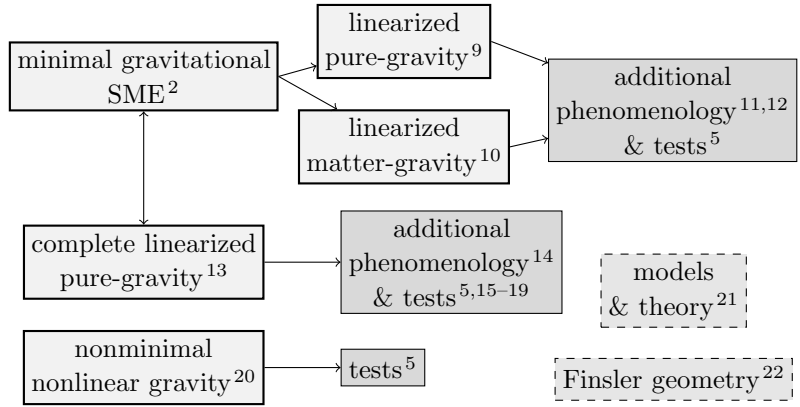

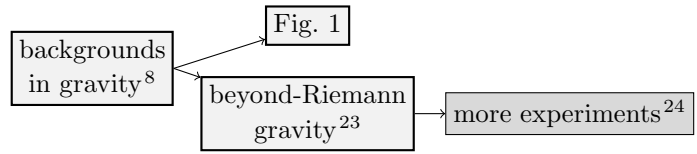
\begin{figure}
\begin{center}
\begin{tikzpicture}[%
    node distance = .4cm,
  inner sep=1mm,
  expt/.style={rectangle, trapezium left angle=120, trapezium right angle=60, thin, fill=gray!30, draw=black, align=center},
  action/.style={rectangle, trapezium left angle=120, trapezium right angle=60, thick, fill=gray!10, draw=black, align=center},
  oact/.style={rectangle, trapezium left angle=120, trapezium right angle=60, dashed, fill=gray!20, draw=black, align=center},
    loop/.style={ellipse, thick, fill=yellow!30, draw=yellow!50, align=center},
  title/.style={font=\LARGE\scshape,node distance=16pt, text=black!40, inner sep=1mm},
  background/.style={rectangle, rounded corners, fill=black!5, draw=black!15, inner sep=4mm}
]

\node[action] (backgrounds)
       {backgrounds \\ in gravity \cite{backgrounds}};
       
 \node[action] (fig1) [above right = -0.2cm and 1.2cm of backgrounds] {Fig.\ 1};
 \node[action] (brg) [below = of fig1] {beyond-Riemann \\ gravity \cite{brg21}};
 
 \node[expt] (brgx) [ right = of brg] {more experiments \cite{brgexpt}};
  
  \draw[->] (backgrounds.east) -- (fig1);
  \draw[->] (backgrounds.east) -- (brg);
  \draw[->] (brg.east) -- (brgx);
  
\end{tikzpicture}
\end{center}
\caption{Additional progress in SME gravity as of CPT'22.
Reference \protect\refcite{backgrounds} can be thought of as a prequel to the work outlined in Fig.\ 1
as well as opening new avenues of investigation.} 
\label{asof22}
\end{figure}

In terms of structure, the SME can be thought of 
as a series expansion about known physics as the level of the action.
The additional terms are
constructed from conventional fields
coupled to coefficients for Lorentz violation,
which can be thought of as providing directionalities to empty spacetime.
The mass dimension of the additional operators labels the order in the expansion.  \cite{jtmd}
The leading terms, which are of mass dimension $d=3,4$,
form a limit known as the minimal SME.
In the gravity sector,
a variety of complementary limits have been explored in the context
of theory, phenomenology, and experiment.
The goal of my contribution to the CPT'19 proceedings was to summarize
the relations among, and the status of, these efforts, \cite{jtcpt19} 
in part through the creation of Fig.\ \ref{asof19}, which remains a useful description of much of the gravity-sector structure,
though additional work has been done in many of its areas.
For additional discussion of Fig.\ \ref{asof19}, see Ref.\ \refcite{jtcpt19}.
The evolution of the field since CPT'19 has led to an understanding of the content of Fig.\ \ref{asof19}
as being but one facet of an expanded array of areas to be explored via the framework of Ref. \refcite{backgrounds}.
Figure \ref{asof22} highlights the addition of this prequel relative to the 2019 structure.

While an effort has been made to point the reader to key works,
those that are recent, 
and those discussed elsewhere in these proceedings,
it is not possible to address all of the work done in this area in this short summary.
We refer the reader to other contributions to these proceedings
along with Refs. \refcite{SME3,datatables} for additional discussion
and references.

\section{Reach, Separation, and Gravitational Waves}
\label{rs}

As can be seen from the data tables,\cite{datatables}
experiments have achieved a high level of sensitivity to coefficients for Lorentz violation
and have explored a large breadth of coefficient space.
In performing such analysis, 
the question of which and how many coefficients to extract measurements for using a given data set naturally arrises.
Practical progress dictates that experimental data be used to extract likelihood bounds on the coefficients for Lorentz violation
in the context of a model involving a subset of the full (and in general infinite) coefficient space of the SME.
This highlights the nature of the SME as a test framework rather than a model.

One popular approach is to consider each SME coefficient one-at-a-time,
perhaps re-using a data set to attain likelihood bounds on multiple coefficients.
This approach is sometimes referred to as a {\it maximum reach} approach\cite{gravimeter}
because it characterizes the maximum reach that the experiment can attain for the coefficient. 
When these measurements are consistent with zero, 
they provide a good order-of-magnitude sense of how big the particular Lorentz-violating effect could be in nature
in the absence of a model involving a fine-tuned cancelation of the effects of multiple coefficients
in the observable under consideration.

When data permits,
it is also common to obtain simultaneous measurements of all or multiple coefficients of the same observer tensor object,
or even several tensor coefficients from a given sector at a given mass dimension.
This is sometimes referred to as a {\it coefficient separation} procedure\cite{gravimeter} and 
it can more definitively exclude a larger set of models.

In my CPT'19 proceedings contribution,\cite{jtcpt19}
I highlighted two key expansions in experimental reach that had recently emerged at that time:
the MICROSCOPE mission in the context of matter-gravity couplings\cite{microscope}
and multimessenger astronomy in the form of gravitational wave (GW) event GW170817
and gamma ray burst GRB 170817A in the context of the minimal gravity sector.\cite{gwgrb}
While little coefficient separation had been done at that time in the context of GW studies,
the now extensive catalog of GW events\cite{gwtc1,gwtc2,gwtc3} has led to a blossoming of these studies.

By taking advantage of the arrival time at the different GW detectors situated around the Earth,
simultaneous measurements of 4 of the 9 minimal gravity-sector coefficients have been achieved\cite{sog} 
using data from the first GW catalog.\cite{gwtc1}
A simultaneous extraction of all 9 minimal gravity sector coefficients using all suitable GW events released to date\cite{gwtc1,gwtc2,gwtc3}
is in preparation.\cite{sog2}

The search for birefringence and dispersion of gravitational waves based on the dimension 5 and 6 coefficients 
has also now been the focus of numerous studies.
Dimension 5 effects have been incorporated into into a version of the Laser Interferometer GW Observatory (LIGO) Algorithm Library suite LALSuite,
and a sensitivity study has been performed using this implementation.\cite{birf}
Results from the body of recent gravitational wave events based on this implementation are in preparation.\cite{oneal2}
An implementation of dimension 5 and 6 effects in the Bilby analysis code has also generated results\cite{niu} based on recent GW events.
Similar work has previously been done based on the duration of LIGO/Virgo chirps.\cite{zwang21}
Additional studies of GW Birefringence that are yet to incorporate direction dependence have also been done,\cite{isobirf}
and isotropic studies of dispersion are ongoing.\cite{haris}

\section*{Acknowledgments}
J.T.\ is supported by NSF grant PHY1806990 to Carleton College
and thanks M.\ Seifert for useful conversations.

\end{document}